# Situational Awareness, Driver's Trust in Automated Driving Systems and Secondary Task Performance

Luke Petersen, Lionel Robert, X. Jessie Yang, and Dawn M. Tilbury

University of Michigan

## Abstract

Driver assistance systems, also called automated driving systems, allow drivers to immerse themselves in non-driving-related tasks. Unfortunately, drivers may not trust the automated driving system, which prevents either handing over the driving task or fully focusing on the secondary task. We assert that enhancing situational awareness can increase a driver's trust in automation. Situational awareness should increase a driver's trust and lead to better secondary task performance. This study manipulated drivers' situational awareness by providing them with different types of information: the control condition provided no information to the driver, the low condition provided a status update, while the high condition provided a status update and a suggested course of action. Data collected included measures of trust, trusting behavior, and task performance through surveys, eye-tracking, and heart rate data. Results show that situational awareness both promoted and moderated the impact of trust in the automated vehicle, leading to better secondary task performance. This result was evident in measures of self-reported trust and trusting behavior.

## 1. Introduction

Driver assistance systems embedded in autonomous and semi-autonomous vehicles have the potential to increase driving safety while providing human drivers with the flexibility to address other pressing issues that they could not address while manually driving (Parasuraman, Cosenzo, and De Visser 2009).

Unfortunately, prior research suggests that not all drivers trust automated driving systems (Beller, Heesen, and Vollrath 2013; Verberne, Ham, and Midden 2012; Xiong et al. 2012). This lack of trust prevents drivers from either handing over driving responsibility or fully focusing on a secondary (i.e. non-driving) task (Gremillion et al. 2016). In the former case, the driver fails to complete the secondary task. In the latter, performance in the secondary task is hindered because the driver is constantly monitoring the driving situation. In order to achieve optimal task performance, drivers must be comfortable relying on the vehicle automation to drive so they can effectively engage in a secondary task. In a civilian setting, typical secondary tasks might include interacting with a navigation device or a smartphone—instances where task engagement might be trivial but can impose sufficiently high attentional demands to make it difficult and unsafe to simultaneously drive the vehicle. In a military setting, typical secondary tasks might include surveillance or mission-critical communications—instances in which mission success is highly dependent on both the driving and non-driving tasks being accomplished properly.

We assert that situational awareness is a significant determinant of trust in automated vehicle capabilities. Trust is promoted by matching an agent's ability to a given situation (Robert, Dennis, and Hung 2009). Trust in the agent (the trustee) occurs when the trustor believes that the trustee's ability is equal to or exceeds the demands of a given task. However, driving is often dynamic and unpredictable, and so to ensure that the situation has not exceeded the vehicle's capability, the driver may be tempted to disengage from the secondary task. A driver assistance system that supports situational awareness should increase a driver's trust in the automated driving system and allow the driver to fully focus on the secondary task, leading to better overall task performance and more likely mission success.

We conducted a human-in-the-loop study with thirty participants. The experimental design consisted of manipulating driver situational awareness and assigning participants a secondary task to complete during a semi-autonomous driving situation. The study manipulated situational awareness in three levels : no situational awareness, low situational awareness, and high situational awareness. The no situational awareness (control) condition provided no information with regard to the driving situation, the low condition provided a status update, while the high condition provided a status update and a suggested course of action.

Our results show that increased situational awareness promotes trust in automation and helps the driver achieve better performance on the secondary task. The results of this study contribute to the literature on trust in automation by providing new insights on the role of situational awareness in driver assistance systems. We discuss several of those contributions and suggest next steps. We believe that future research could build on our findings to further expand our knowledge on situational awareness in driver assistance systems and trust in automation.

The remainder of this paper is organized as follows. Section 2 reviews related work, provides additional background and motivation, and presents the user study's expected outcomes. Section 3 describes the design of the user study. Section 4 presents results from the user study, and Section 5 discusses these results.  Section 6 presents some limitations and conclusions from the paper.

## 2. Background and Expected Outcomes

In this section, we discuss measuring trust in automation and evaluating situational awareness. We compare our study against related studies in the literature and discuss the expected study outcomes.

### 2.1 Trust in Automation

Most definitions of trust in automation involve expectations, confidence, risk, uncertainty, reliance, and vulnerability (Billings, Schaefer, Llorens, & Hancock, 2012). No single definition exists for trust in Automated Vehicles, or AVs (Saleh, Hossny, & Nahavandi, 2017). In this paper, we draw from two popular definitions of trust by Mayer, Davis, and Schoorman (1995) and Lee and See (2004). Mayer et al. (1995) defined trust as "the willingness of a party to be vulnerable to the actions of another…, irrespective of the ability to monitor or control that other party." Lee and See (2004) defined trust in automation as "the attitude that an agent will help achieve an individual's goals in a situation characterized by uncertainty and vulnerability." For our research, we leverage both definitions and define trust in AVs as the willingness of an individual to be vulnerable to the actions of an AV based on the attitude that the AV will help them achieve their goals. Trust in an autonomous vehicle has been identified as a vital determinant of whether the driver will employ an autonomous vehicle (Beller, Heesen, and Vollrath 2013; Charalampous, Kostavelis, and Gasteratos 2017; Verberne, Ham, and Midden 2012; Xiong et al. 2012).

Previous studies examining approaches to facilitating trust in autonomous or semi-autonomous vehicles were designed around a common paradigm that assumes that the driver is or should be monitoring the vehicle actions at all times to take over if needed (Carsten et al. 2012; Hergeth et al. 2016; Ma and Kaber 2005). As a result, these studies often treat secondary (i.e. non-driving) tasks as distractions or challenges to compensate for or prevent, rather than to support. This view is appropriate when the goal is to promote the driver's attention toward the driving. However, this view ignores the potential of drivers to fully leverage the benefits of autonomous or semi-autonomous driving by intentionally not focusing on the vehicle's driving all the time. In these situations, secondary tasks are not distractions and should be supported.

In our study we measured trust in several ways. Traditionally, trust has been captured via subjective self-reported rating. We employed one of the most widely accepted self-reported measures (Muir and Moray 1996). We have also included more objective behavioral and physiological measures.

Behavioral measures associated with trust in our study included driver monitoring via eye gaze and compliance with the driver assistance systems. Monitoring has normally been associated with a lack of trust in many settings (Robert 2016). Monitoring of automation has been negatively associated with trust in automation (Hergeth et al. 2016). Measures of compliance include distance to impact with another vehicle and waiting time before taking over control of the driving. We expect drivers who trust the automated driving system to wait for voice commands from the driver assistance systems before acting. This would mean that the trusting driver would wait longer before taking control over the driving and therefore come closer to hitting another vehicle.

We also included measures of heart rate variability and heart rate (beats per minute, or bpm). Heart rate variability has been used as an index of perceptions of threat to one's safety and stress (Thayer and Lane 2007). Generally, lower heart rate variability has been associated with greater perceptions of threat to safety and stress (Thayer et al. 2012). Therefore, we expected heart rate variability to be positively related to trust in automation. Increases in the heart rate (beats per minute) are associated with increases in stress (Meehan et al. 2002). When individuals trust the vehicle, they should be less worried and have less concern. Therefore, increases in trust in automation should coincide with decreases in heart rate.

*2.2 Situational Awareness*

Situational awareness is defined as the perception and comprehension of information that allows an individual to project future courses of action needed to respond to a dynamic environment (Endsley 1995a). Recently, researchers have suggested that situational awareness might help promote trust in automated driving by allowing the driver to better understand the environment and predict what future actions, if any, are needed (Miller, Sun, and Ju 2014). We build on this prior literature on situational awareness and trust in automation by asserting that driver assistance systems that promote situational awareness should increase trust in automation and ultimately result in better secondary task performance by allowing the driver to focus more on the secondary task.

Despite the potential of situational awareness to support the driver's performance on a secondary task, little if any research has been directed in this area. Although the research in situational awareness is plentiful in the human–robot interaction (HRI) domain, substantially fewer studies have been conducted in the context of autonomous (or semi-autonomous) driving. Existing studies have looked primarily at two areas: graphically displaying some sort of "uncertainty" metric to the driver (Beller, Heesen, and Vollrath 2013; Helldin et al. 2013; Rezvani et al. 2016; Stockert, Richardson, and Lienkamp 2015), and looking at the effect of missed alarms and false alarms on a driver's response to feedback from the vehicle (Lees and Lee 2007; Maltz and Shinar 2004). We addressed this research gap in our study and contribute to the field of situational awareness research by directly examining the relationship among situational awareness, trust in automated driving systems, and secondary task performance.

Several common instruments are used to measure situational awareness. The two most common are the situation awareness global assessment technique (SAGAT), taken from Endsley (1995b), and the situation awareness rating technique (SART), taken from Taylor (1990). SAGAT is a freeze-on-line probe that requires researchers to pause and intervene during the task to make queries about elements in the situation. Queries are aligned with the three constructs of situational awareness in Endsley's model of situational awareness (Endsley 1995a). SAGAT minimizes attention bias because subjects cannot anticipate the queries in advance; however, it can be intrusive because it requires pausing the task at critical times of task engagement. Therefore, we employed SART, which is administered to participants after they perform a task, to determine participant situational awareness. The items measured through the SART use a seven-point rating scale (1=low, 7=high), which asks participants to rate their experience in the performed task.

SART measures situational awareness via three dimensions—demands on attentional resources (SART-DAR), supply of attentional resources (SART-SAR), and understanding of the situation (SART-UOS)—and uses all three to create an overall measure of situational awareness (Selcon and Taylor 1990; Taylor 1990). Table 1 shows the list of abbreviations pertinent to the SA measures.

*2.3 Expected Outcomes*

Situational awareness in the context of an automated vehicle can refer to the degree to which the driver is aware of the current and future driving conditions facing the vehicle. Situational awareness should be particularly important when a driver's secondary task requires them to disengage from the driving. The driver must trust the automated vehicle while focusing on performing a secondary task. Performance on the driver's secondary task is hindered when the driver refuses to trust the automated vehicle and attempts to engage in the driving and secondary tasks simultaneously. When the automated vehicle supports situational awareness through driver assistance systems, the driver can forecast potential problems and take control before these events occur. Under these conditions, drivers are more likely to trust the automated vehicle and focus fully on the secondary task. Therefore, the basic premise of our research is this: As the level of situational awareness increases, so should the trust in automation.

In our study, we varied situational awareness by changing the content of a verbal message presented to the driver. The message provided information relative to an upcoming obstacle on the roadway. We varied the degree of situational awareness support across three conditions. In the control condition, no message was provided. In the low situational awareness condition, a simple status update was provided. In the high situational awareness condition, the same simple status update was presented followed by a suggested course of action. Based on the prior literature on situational awareness and trust, we have the following expected outcomes:

EO1. As we move from the control condition to the high situational awareness condition, situational awareness will increase.
EO2. As we move from the control condition to the high situational awareness condition, self-reported trust will increase.
EO3. As we move from the control condition to the high situational awareness condition, behavioral and physiological measures of trust will increase.
EO4. As we move from the control condition to the high situational awareness condition, the impact of trust on secondary task performance should increase.

**Table 1: A list of abbreviations**

| Abbreviation | Terminology |
|---|---|
| SAGAT | situation awareness global assessment technique |
| SART | situation awareness rating technique |
| SART-DAR | situation awareness rating technique -- demands on attentional resources |
| SART-SAR | situation awareness rating technique -- supply of attentional resources |
| SART-UOS | situation awareness rating technique -- understanding of the situation |

**Table 2  Participant characteristics.** A total of thirty-three drivers with valid U.S. driver's licenses voluntarily participated in the study. Three participants were excluded from analysis because of simulation errors during the experiment.

| Gender | 8 females | 22 males |
|---|---|---|
| Handedness | 29 right-handed | 1 left-handed |

| | Mean | StDev |
|---|---|---|
| Age | 25.7 years | 5.5 years |
| Driving experience | 7.8 years | 5.7 years |

**Table 3** Self-reported usage of driver assistance features. This survey was administered immediately following consent and prior to the explanation of the experiment. Most subjects had interacted with cruise control; few had utilized higher-level assistance features. Percentages may not add to 100% due to rounding.

| | Frequency of Use of Driver Assistance Features | | | | |
|---|---|---|---|---|---|
| | **Never** | **Once** | **Periodically** | **Frequently** | **Always** |
| **Driver Assistance Feature** | | | | | |
| **Cruise control** | 10% | 7% | **40%** | **33%** | **10%** |
| Adaptive cruise control | 80% | 10% | 3% | 3% | 3% |
| Lane departure warning | 80% | 7% | 10% | 0% | 3% |
| Lane-keeping assistance | 87% | 3% | 7% | 0% | 3% |
| Forward collision warning | 73% | 17% | 3% | 3% | 3% |

## 3. User Study

In this research we utilized a human-in-the-loop study to evaluate driver trust in automation when the automated driving system purposefully augmented the driver's situational awareness.

### 3.1 Participants

A total of thirty-three drivers with valid U.S. driver's licenses voluntarily participated in the study. Participants were recruited from a Midwestern public university. Three participants were excluded from analysis because of simulation errors during the experiment. The remaining thirty participants consisted of twenty-two males and eight females with an average age of 25.7 years (standard deviation [SD]=5.5 years) and an average driving experience of 7.8 years (SD=5.7 years). Twenty-nine reported right-handedness, while one reported left-handedness; this information is summarized in Table 2. Prior research has not reported consistent gender differences on trust in automation (Hoff & Bashir, 2015).

The participants also reported their experience with various driver assistance features. Their responses are tabulated in Table 3.

### 3.2 Procedure

Participants first completed a consent form to participate in the study. Next, participants completed a pre-experiment survey. The pre-experiment survey consisted of questions about demographic information as well as experience using driving aids, such as adaptive cruise control and forward collision warning. It also included questions to determine each participant's propensity to trust automation, derived from Singh, Molloy, and Parasuraman (1993).

After completing the pre-experiment survey, participants completed a brief training session to become familiar with the vehicle controls and the (non-driving) secondary task. Following training, the eye-tracker and heart rate monitor were fitted and calibrated. Participants then completed three test sessions, one corresponding to each of the situational awareness conditions (described in detail in the next section). Each driving session lasted approximately 10 minutes. At the end of each session, participants completed the post-condition survey. The post-condition survey included measures for SART taken from Taylor (1990) as well as questions about trust in automation, derived from Muir and Moray (1996) and adapted to suit this study. All surveys were administered via web-form. Each experiment lasted approximately 90 minutes.

*3.3 Tasks*

Participants were tasked with operating a simulated semi-automated vehicle while also attending to a visually engaging secondary task. The simulated vehicle was equipped with lane-keeping, speed-maintenance, and automatic emergency braking capabilities. Additionally, the simulated vehicle delivered auditory messages regarding obstacles on the road. The content of these messages was varied across the three driving sessions and is discussed in detail in the "Study Design" section.

The virtual driving scenario was a standard two-lane highway with a hard shoulder and a posted speed limit of 70 mph. In lieu of a dashboard, a heads-up display overlaid the vehicle speed and driving mode (manual or automated) on the driving view (see Figure 1). Participants were told that the simulated vehicle could drive itself, but that given the highway speeds it would not be able to maneuver around a stopped obstacle obstructing the roadway. In these circumstances, participants would have to take over control of the vehicle by turning the steering wheel or applying the brake. Failure to do so would result in the simulated vehicle automatically emergency braking. No other moving traffic was present in the scenario; however, at certain points in the simulation, stopped vehicles would appear on the road ahead. Participants encountered four such stopped vehicles in each of the driving sessions. In each session, either two or three of the stopped vehicles would appear in the same lane as the driven vehicle, thus requiring action by the participant. The other stopped vehicle(s) appeared in the opposite lane.

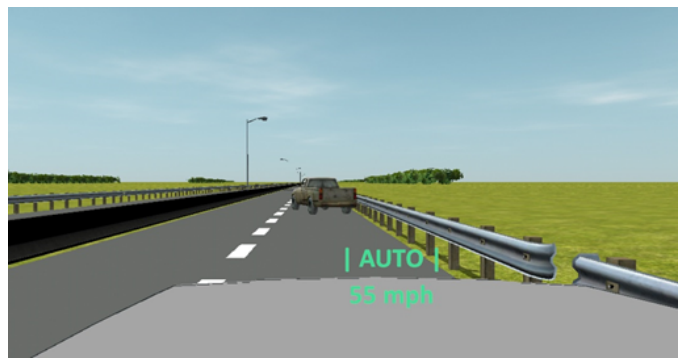

**Figure 1** Simulated driving view, rendered with ANVEL software. Vehicle speed and driving mode are displayed in a heads-up display (HUD)

The secondary task (see Figure 2) was a modified version of the surrogate reference task (International Organization for Standardization 2012). The surrogate reference task resembles a target recognition task, in which participants are required to identify a target item (the letter Q in this study) amid a field of distractors (the letter O) and manually select it on a touchscreen located to the right of the participant. This secondary task is commonly used in studies of this nature (Beller, Heesen, and Vollrath 2013; Hergeth et al. 2016; Hsieh, Seaman, and Young 2015; Petermeijer et al. 2017; Stockert, Richardson, and Lienkamp 2015). The task imposes a controllable level of cognitive load and resembles an ordinary activity like interacting with an infotainment system or smartphone.

During the study, participants competed for monetary bonuses. Participants earned one point for every completed secondary task. Participants were penalized fifteen points if the driven vehicle got within 250 feet (76.2 meters) of a stopped vehicle before switching lanes. Participants were penalized twenty-five points if a collision with a stopped vehicle occurred or last-minute automatic braking was induced. The highest-scoring drivers received cash bonuses. This scoring structure forced participants to consider their willingness to rely on the vehicle's driver assistance system so they could focus on the secondary task.

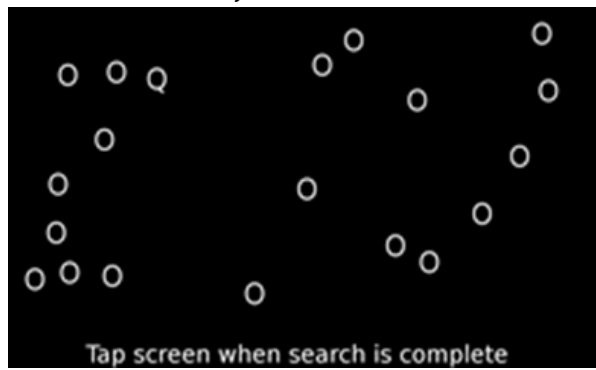

**Figure 2** Example view of secondary task. This task was administered on a touch screen and required subjects to manually select the target shape (the letter 'Q', in the upper left quadrant)

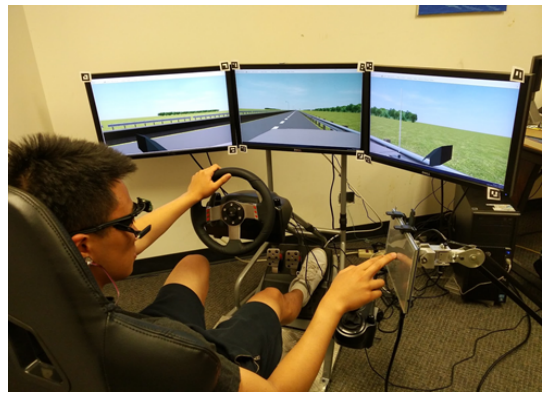

**Figure 3** Driving simulator and secondary task setup. This setup increased the realism of the experimental tasks. Markers were placed on each monitor and the touchscreen to identify surfaces for eye-tracking

*3.5 Study Design*

The study employed a one-way within-subjects design. The single independent variable in this experiment was the content of auditory message presented to the driver. Each participant performed the experiment under all three conditions of this messaging, with each condition corresponding to a degree of situational awareness support. The presentation order of the three conditions was counterbalanced using a Latin square design to minimize learning and ordering effects. The control condition is considered as a baseline wherein no automation in available (Wickens et al., 2010). These three conditions are tabulated in Table 4. During each driving session, the participant encountered four stopped vehicles on the roadway. As discussed, only a subset of these stopped vehicles appeared in the same lane as the driven vehicle and thus required driver intervention to prevent a collision. In the low situational awareness condition, the message was played 5 seconds (512 feet; 156 meters) prior to reaching the stopped vehicle. In the high situational awareness condition, the first message was played 6.5 seconds (656 feet; 200 meters) prior, and the follow-up message was played at 5 seconds. Timelines for each of the conditions can be seen in Figure 4. Only the sequences for the first two stopped cars are shown. This pattern of sequences was repeated for each stopped car during the simulation.

**Table 4** Conditions of the independent variable: auditory message content

| Condition | Auditory Message | Circumstance |
| --- | --- | --- |
| Control | None | |
| Low SA | "Stopped vehicle ahead" | For all stopped vehicles |
| High SA | "Stopped vehicle ahead" followed by | For all stopped vehicles |
| | "No action needed" | For stopped vehicles in a lane other than that of the driven vehicle |
| | OR | |
| | "Take control now" | For stopped vehicles in the same lane as the driven vehicle |

*SA* situational awareness

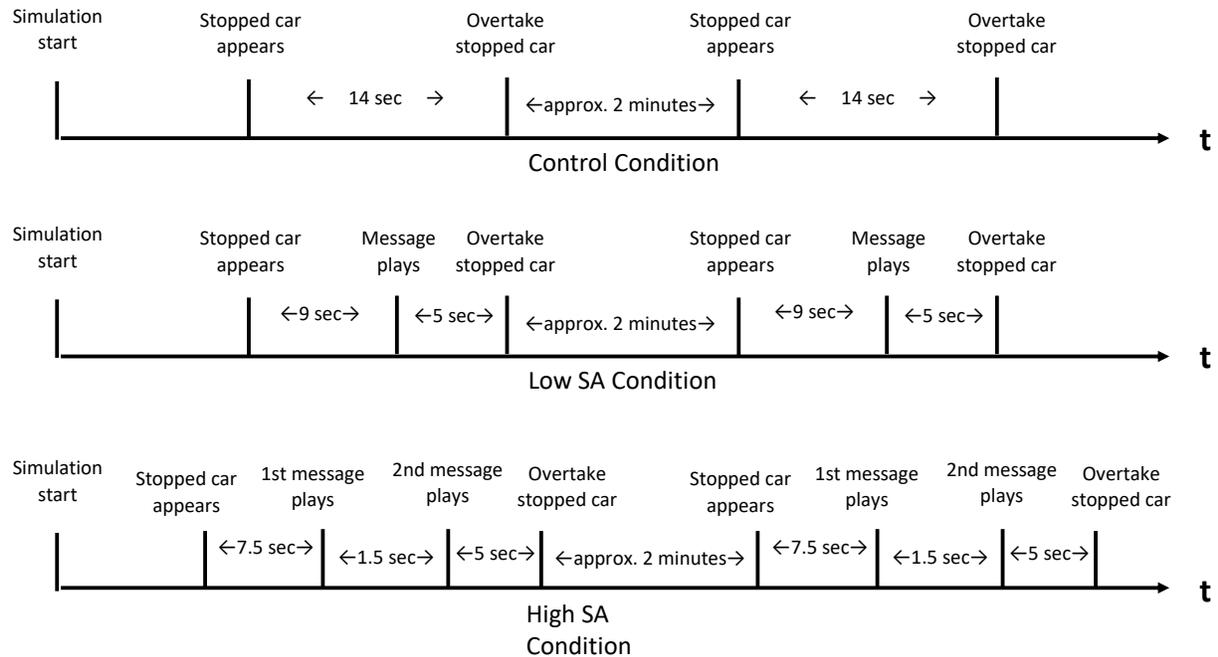

**Figure 4** Driving session timelines. During each session, the subject encountered four stopped vehicles. The timing of messages, corresponding to the situational awareness condition, were relative to the appearance of each stopped vehicle and the calculated time-to-collision (i.e. the "Overtake stopped car" marks). The timelines for the first two stopped cars repeated accordingly for the subsequent two stopped cars. *SA* situational awareness

*3.6 Measures*

We collected the following dependent measures:

1. Eye-tracking data: monitoring ratio (proportion of time spent looking at the driving scene) and monitoring frequency (rate at which visual attention is switched between areas of interest).

2. Heart rate data: number of heart beats per minute (bpm) and heart rate variability (HRV, computed as Root Mean Square of the Successive Differences, rMSSD), measured with a heart rate monitor.

3. Driving data: simulated vehicle state (position; heading; velocity; yaw rate; and acceleration); proximity to the nearest upcoming stopped vehicle; participant take-over behavior, including steering input and pedal input;

4. Participant secondary task engagement, including whether selection is correct, selection time, and total number of correct selections.

5. Survey responses:
    a. Pre-experiment:
    i. Demographic and driving experience
    ii. Propensity to trust automation (Singh, Molloy, and Parasuraman 1993)
    b. After each condition:
    i. Self-reported trust via trust in automation survey (Muir and Moray 1996)
    ii. Situation awareness via situation awareness rating technique (SART) (Taylor 1990)

*3.7 Apparatus*

We conducted the study using a static driving simulator with three screens. We used Autonomous Navigation Virtual Environment Laboratory (Durst et al. 2012) to create the virtual environment and implement the automated driving behavior. The secondary task was administered on a 10.1-inch touchscreen mounted to the right of the driver, in a position representative of where a vehicle's center console would be in an actual vehicle. We used PEBL (Psychology Experiment Building Language) (Mueller and Piper 2014) to create this secondary task. A head-mounted eye-tracker collected participant gaze activity during the study. This device captures video of the wearer's field of view and of the wearer's right eye. We used paper markers in conjunction with this software to define surfaces of interest, namely the three simulator monitors and the touchscreen. We also monitored heart rate and heart rate variability during the study.

4. Results

We employed SPSS (Statistical Package for the Social Sciences) version 24 mixed linear model package to conduct all our analysis. Mixed linear models are statistical models that contain both fixed and random effects in their estimations (West, Welch, & Galecki, 2007). By modeling both fixed and random effects, they can accommodate data which is correlated or non-independent. Our data consisted of three conditions per participant. Therefore, each of the observations per condition was nested within an individual, violating one of the assumptions of ordinary least squares (OLS) regarding the independence of observation. Mixed linear models can account for the lack of independence. Table 5 contains the estimated marginal means and standard errors of all the analyzed measures. In the next sections, we discuss the analysis of each measure in greater detail in relation to its respective expected outcomes.

**Table 5** Estimated marginal means and standard errors using mixed linear modeling; measures in bold are significant.

| Category | Measure | Control | Low SA | High SA | Sig. |
|---|---|---|---|---|---|
| Situational awareness (SA) | Situational awareness: demand on attentional resources (SART-DAR) | 2.8 (SE = .23) | 2.3 (SE = 0.16) | 2.2 (SE = 0.19) | p = 0.06 |
| | Situational awareness: supply of attentional resources (SART-SAR) | 3.6 (SE = 0.21) | 3.9 (SE = 0.20) | 2.2 (SD = 0.23) | p = 0.07 |
| | **Situational awareness: understanding (SART-UOS)** | 4.3 (SE = 0.27) | 4.3 (SE = 0.24) | 4.7 (SE = 0.23) | **p = 0.03*** |
| | **Situational awareness: overall** | 7.9 (SE = 0.32) | 8.2 (SE = 0.33) | 8.9 (SE = 0.32) | **p = 0.00**** |
| Trust | **Self-reported trust** | 27.8 (SE = 1.2) | 30.6 (SE = 0.55) | 31.3 (SE = 0.68) | **p = 0.02*** |
| Behavioral | **Time to take control after stopped car appears (ms)** | 9,100 (SE = 510) | 8,000 (SE = 590) | 9,900 (SE = 500) | **p = 0.04*** |
| | **In-lane distance from stopped car before lane change (meters)** | 97 (SE = 12) | 130 (SE = 14) | 85 (SE = 11) | **p = 0.048*** |
| Physiological | **Monitoring ratio (proportion of time looking on-road)** | 0.180 (SE = 0.018) | 0.120 (SE = 0.013) | 0.125 (SE = 0.014) | **p = 0.025*** |
| | Monitoring frequency (Hz; frequency of switching fixation between on-road and off-road) | 0.44 (SE = 0.05) | 0.36 (SE = 0.05) | 0.36 (SE = 0.04) | p = 0.41 |
| | Heart rate (BPM) | 110 (SE = 9.2) | 110 (SE = 9.3) | 120 (SE = 9.07) | p = 0.81 |
| | Heart rate variability (ms) | 650 (SE = 43) | 590 (SE = 43) | 620 (SE = 46) | p = 0.63 |

*SD* standard deviation, *SE* standard error. $P<.05$ was significant

**EO1: As we move from the control condition to the high situational awareness condition, situational awareness will increase.**

We found significant differences among the three conditions (see Table 5) with regard to SART-UOS (p=.03), less so with SART-DAR (p=.06) and SART-SAR (p=.07). We found no differences between the high situational awareness and low situational awareness conditions with regard to SART-DAR (p>.05) or SART-SAR (p>.05). However, both conditions were significantly different from the control condition with regard to SART-DAR (p<.05) and SART-SAR (p<.05). We found significant differences with regard to SART-UOS between the high condition and both the low (p<.05) and control (p<.05) conditions. There were no differences between the low and control conditions with regard to SART-UOS (p>.05). In summary, both levels of situational awareness seem to reduce demands on attentional resource (SART-DAR) and increase supply of attentional resources (SART-SAR) when compared to the control condition. But, only the high situational awareness condition led to increases in the understanding of the situation that were significant (p<.05).

We then calculated an overall score of situational awareness based on similar procedures recommended by both Selcon and Taylor (1990) and Taylor (1990). We calculated a composite SART score using the means of each sub-dimension following this formula: situational awareness = SART-UOS + (SART-SAR – SART-DAR). When the means were compared, there were clear differences among the three conditions (p<.01). However, only the high situational awareness condition was significantly different from the other conditions (p<.05). There was no significant difference between the low situational awareness and control situational awareness conditions. Overall, it appears that only the high situational awareness condition leads to a significantly higher level of situational awareness as measured by SART.

**EO2: As we move from the control condition to the high situational awareness condition, self-reported trust will increase.**

Next, we examined the impact of the conditions on trust in the automated vehicle. The reliability of trust in automation was $\alpha = .86$, well above the .70 requirement (Nunnally 1978). As seen in Table 5, the high situational awareness condition had the highest mean of self-reported trust (31.3), followed by the low situational awareness condition (30.6), then the control condition (27.8). There were no differences between the high and low conditions (p>.05); however, the high and low conditions were both significantly different from the control condition (p<.05). In sum, the low and high situational awareness conditions led to higher trust in the automated vehicle than the control condition.

**EO3a: As we move from the control condition to the high situational awareness condition, behavioral measures of trust will increase.**

We then examined trust in the form of behavioral measures. We used measures related to driver gaze behavior: (1a) monitoring ratio and (1b) monitoring frequency. "Monitoring" refers to the extent that the driver glanced at the driving scene. Surface trackers were located on the three computer monitors and the touchscreen. Glances, defined as a fixation of 120 milliseconds [ms] or longer (Graf & Krebs, 1976), at the monitors were considered "on-road," whereas all other glances were considered "off-road." The eye-tracking metrics were benchmarked against prior literature (Hergeth et al. 2016). Monitoring ratio was measured as the proportion of time drivers spent looking at the road, while monitoring frequency was measured as the rate at which drivers switched their view between on-road and off-road.

The results showed a significant difference in monitoring ratio among the three conditions (see Table 5). Additional analysis showed significant differences at the .05 level between the control condition (.180; standard error [SE]=0.018) and both the low situational awareness condition (.120; SE=0.013) and the high situational awareness condition (.125; SE=0.014). However, there were no differences between the low and high situational awareness conditions. Differences in monitoring frequency were not significant across the three conditions.

Next, we examined two behavioral measures related to driving. The measures were operationalized in the form of two related variables: (1) the mean time participants waited before taking over control of the vehicle after the stopped car appeared and (2) the minimum distance between the driven vehicle and the stopped car before the drivers changed lanes. Larger values of (1) and small values of (2) are indications that participants trusted the vehicle enough to wait longer before actually taking control and changing lanes. Results show that participants waited longer to take over control after the car appeared in the high situational awareness condition (mean=9,900 ms; SE=500 ms) compared to the situational awareness low condition (mean=8,000 ms; SE=590 ms). The difference was significant at p<.05. There were no such differences involving the high situational awareness condition and the control condition (mean=9,100 ms; SE=510 ms). Note that in the low situational awareness condition, the time between each stopped car appearing and the message playing was about 9 seconds. In the high situational awareness condition, this difference was about 7.5 seconds.

Similarly, participants stayed in the lane longer before switching lanes in the high situational awareness condition (mean=85 meters [m]; SE=11) compared to the low situational awareness condition (mean=129 m; SE=13). The difference was significant at p<.05. There were no such differences between the high situational awareness condition and the control condition (mean=97 m; SE=12). In sum, in the high situational awareness condition participants waited longer before taking control over the vehicle after the stopped car appeared and got closer to the stopped car before switching lanes than in the low condition. However, there were no such differences with respect to the control condition in either case.

**EO3b: As we move from the control condition to the high situational awareness condition, physiological measures of trust will increase.**

We evaluated two physiological measures of trust. The first two measures related to heart rate: (2a) heart rate (bpm) and (2b) heart rate variability (ms). As shown in Table 5, differences in heart rate and heart rate variability among the three conditions were not significant.

**EO4: As we move from the control condition to the high situational awareness condition, the impact of trust on secondary task performance should increase.**

We developed three models with performance on the secondary task as the dependent variable. Secondary task performance was defined as the total number of search tasks the driver correctly completed during the driving simulation. Our goal was to examine if the impact of SA on secondary task performance was dependent on driver's trust in the autonomy. To accomplish this, we needed to demonstrate that not only was the moderation effect significant but also that its inclusion into our model provided a significant increase in our variance explained. Our first step was to assess how much of the variance was due to factors outside of our experimental variables. Model 1 only included demographic or control variables: hand preference, age, experience, and propensity to trust, which were not manipulated. Model 2 allowed us to assess the variance explained by adding in our experimental variables (i.e. trust in the automated vehicle and SA) along with the control variables. Model 3 included the moderation which we could compare against Model 2 to determine if the additional variance explained by the inclusion of the moderation was significant over and above the impacts of the control and experimental variables. All three models were linear as none included any polynomial variables. The results of model 1, model 2 and model 3 can be seen in Table 6, Table 7, and Table 8, respectively.

**Table 6** Model 1 with only control variables. None of the control variables are significant.

| Parameter | Estimate | SE | df | T | Sig. | Lower Bound | Upper Bound |
|---|---|---|---|---|---|---|---|
| Intercept | 145.21*** | 2.94 | 22.97 | 49.31 | 0.00 | 139.12 | 151.31 |
| Hand Preference | 13.37 | 16.42 | 22.97 | 0.81 | 0.42 | -47.34 | 20.61 |
| Age | -10.32 | 6.62 | 22.97 | -1.56 | 0.13 | -24.00 | 3.37 |
| Driver Experience | 4.13 | 6.63 | 22.97 | 0.62 | 0.54 | -9.58 | 17.85 |
| Propensity to Trust | 1.57 | 3.01 | 22.97 | 0.52 | 0.61 | -4.65 | 7.80 |
| Dependent Variable: Performance on Secondary Task ||||||||
| *p<.05; **p<.01; ***p<.001 ||||||||

**Table 7** Model 2 with control variables and independent variables. Neither of the independent variables are significant.

| Parameter | Estimate | SE | df | T | Sig. | Lower Bound | Upper Bound |
|---|---|---|---|---|---|---|---|
| Intercept | 144.75*** | 3.06 | 22.30 | | 0.00 | 138.42 | 151.09 |
| Hand Preference | 10.82 | 17.02 | 23.33 | 47.33 | 0.53 | -46.00 | 24.36 |
| Age | -9.60 | 6.78 | 22.67 | 0.00 | 0.17 | -23.63 | 4.44 |
| Driver Experience | 3.54 | 6.80 | 22.74 | -0.64 | 0.61 | -10.54 | 17.62 |
| Propensity to Trust | 0.77 | 3.16 | 24.70 | 0.53 | 0.81 | -5.74 | 7.28 |
| SA Condition (SA) | 0.32 | 1.31 | 27.47 | -1.42 | 0.81 | -2.36 | 3.00 |
| Trust in Autonomy (TA) | 2.80 | 2.03 | 49.96 | 0.17 | 0.17 | -1.28 | 6.88 |
| Dependent Variable: Performance on Secondary Task ||||||||
| *p<.05; **p<.01; ***p<.001 ||||||||

**Table 8** Model 3 with control variables, independent variables and moderation effect. The moderation effect of the two independent variables (situational awareness and trust in autonomy) is significant in predicting the performance of the secondary task.

| Parameter | Estimate | SE | df | T | Sig. | Lower Bound | Upper Bound |
|---|---|---|---|---|---|---|---|
| Intercept | 143.16*** | 3.06 | 23.441 | 46.718 | 0.00 | 136.83 | 149.49 |
| Hand Preference | -3.37 | 17 | 24.12 | -0.198 | 0.85 | -38.45 | 31.72 |
| Age | -8.13 | 6.69 | 22.836 | -1.215 | 0.24 | -21.97 | 5.72 |
| Driver Experience | 2.29 | 6.71 | 22.918 | 0.342 | 0.74 | -11.59 | 16.18 |
| Propensity to Trust | -0.91 | 3.16 | 25.424 | -0.289 | 0.78 | -7.41 | 5.58 |
| SA Condition (SA) | -0.29 | 1.19 | 27.286 | -0.248 | 0.81 | -2.72 | 2.14 |
| Trust in Autonomy (TA) | **5.71*** | 2.08 | 49.279 | 2.749 | 0.00 | 1.54 | 9.88 |
| TA*SA | **3.83*** | 1.27 | 27.88 | 3.019 | 0.00 | 1.23 | 6.43 |
| Dependent Variable: Performance on Secondary Task ||||||||
| *p<.05; **p<.01; ***p<.001 ||||||||

As shown in Table 8, the situational awareness level moderated the impact on trust in the autonomy on secondary task performance ($\beta$ =3.83, p<.001). Figure 5 shows how secondary task performance was affected by the situational awareness condition and trust in the autonomy. Low trust is represented on the x-axis by one standard deviation below the mean, while high trust is represented by one standard deviation above the mean. As self-reported trust in the automation increases, its relationship with performance on the secondary task varies by the situational awareness condition. In the high situational awareness condition, trust had a strong positive impact on secondary task performance. However, when there was no or low situational awareness (i.e. in the control condition, or low situational awareness), trust had little impact on secondary task performance. In sum, Figure 5 clearly shows that the effects of trust in automation on the performance of the secondary task are dependent on the situational awareness condition.

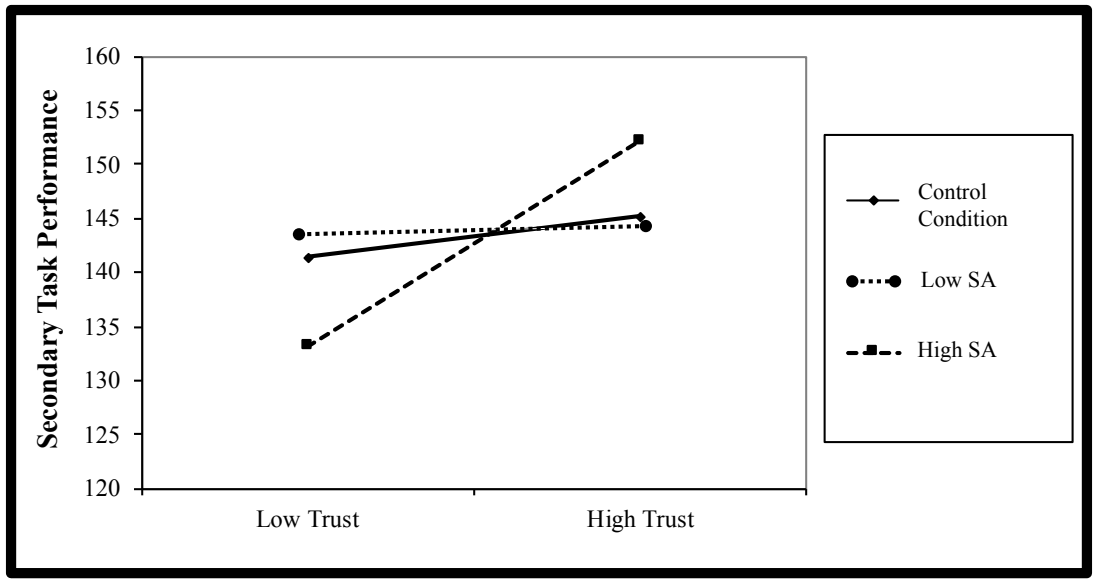

**Figure 5** Moderation of trust by the condition on the secondary task performance. We measured secondary task performance by the number of touchscreen tasks completed during the duration of the driving session. This figure shows that trust had a more significant influence on performance in the high situational awareness (SA) condition as compared to the control condition. This effect was not observed in the low situational awareness condition

Our results can be organized around four overarching findings. One, we found that only the high situational awareness condition led to a significantly higher level of measured situational awareness. This difference in situational awareness seems to be in large part because of the differences in the "understanding of the situation" component of situational awareness as measured by the SART. That is, the low situational awareness condition was found to be inadequate to improve the drivers' comprehension of the obstacles on the roadway. Two, the use of either the high or low situational awareness condition led to increases in trust in automation over the control condition. Three, both the high situational awareness and low situational awareness conditions led to more trusting behaviors as measured by monitoring ratio but not monitoring frequency, but the high situational awareness condition led to significantly higher levels of trusting behavior as reflected in driver behavior when compared to the low situational awareness condition. Unexpectedly, participants showed the least trusting behavior in the low situational awareness condition and not the control condition. Finally, the impact of trust on the performance of the secondary task depended on the level of situational awareness provided by the vehicle.

## 5. Discussion

Overall, the results of this study contribute to the literature on trust in automation, particularized to automated vehicles, in several ways. First, this study demonstrated the importance of driver assistance systems in supporting situational awareness to facilitate trust in automation. The ability of drivers to be aware of the driving situation is vital to encouraging effective use of automated vehicles. We provide evidence that driver assistance systems that support situational awareness can have a positive impact on drivers' trust in automated vehicles.

Second, this research enhances our understanding of the importance of the level of situational awareness supported by driver assistance systems. We show that driver assistance systems need to include not only status updates but projection about future events (the high situational awareness condition) to improve the driver's understanding of the situation. We also show that there are times when projection (high situational awareness condition) has no clear benefits over providing the current status (low situational awareness condition). For example, as discussed, there were no significant differences in self-reported trust between the low situational awareness condition and the high situational awareness conditions. This finding is important because projection (the high situational awareness condition) might require the use of additional computational power. It might be important to know under what circumstances additional computational power is warranted, to help understand how to design better user interfaces to support situational awareness (see Alcázar, Martinez, Pantoja, Collazos, and Pa, 2014).

Although we expected differences between the low and high SA conditions, they were not always significant. There are at least two explanations. One explanation is that the benefits of projection (i.e. high SA) were not always important relative to the benefits of just providing status updates (i.e. low SA). In other words, letting participants know that a vehicle was ahead was enough. Another explanation is that participants may have thought that the information provided by the low SA condition coupled with their own judgement was sufficient. This may explain why there were no significant differences with regards to reported trust, but, clear differences with regard to behavioral measures. One way to explore this line of inquiry is to conduct future research which varies the difficulty of the task. The differences between low and high SA may be more apparent as the degree of task difficulty increases.

Third, surprisingly we found a disconnect between the self-reported trust and observed trusting behaviors. There was no difference between the high and low situational awareness conditions with regard to trust in automation as measured by the self-reported survey. However, clear differences were observed between the two conditions with regard to trusting behaviors: time and distance to take control over the automated vehicle. In the high situational awareness condition, participants waited longer and allowed the vehicle to come closer to the upcoming stopped vehicle before taking control, indicating a higher level of trust. Increases in waiting allowed participants to focus more on the secondary task and earn higher scores. Thus, the benefits of projection might not show up by asking participants how much they trust the automated vehicle, but rather by observing their driving behavior. Future research could include a deeper investigation in the relationship between trust and trusting behaviors.

We noticed a disconnect between self-reported trust and observed trusting behaviors in our study. This study is not the first to discover such differences. Scholars have found both strong (Ho, Wheatley & Scialfa, 2005) and weak relationships (McBride et al., 2010) between trust and trusting behaviors. In fact, there has long been discussions in the field of psychology dedicated to explaining why attitudes are often not related to their corresponding behaviors (see Wilson et al., 1989). In our study, we found evidence of both a strong (i.e. monitoring ratio) and a weak (i.e. take over time, distance to collision, HRV measures) relationship between self-reported trust and trusting behaviors. One explanation for our weak relationships with regards to take over time and distance could be a disconnection between the participant's attitude regarding trust and those measures. More specifically, we assumed that participants who trusted the ADS more would take over driving later at a much closer distance to collision than those who did not trust the ADS. However, for at least some participants, trusting the ADS may have had little to do with how long they waited before taking over the driving. An explanation for the non-significant relationship between HRV and self-reported trust could be due to the task complexity involved in our study. We assumed that participants who trusted the ADS more would have significantly lower HRV measures. However, regardless of the level of trust, participants were actively engaged in a complex task which may have kept their heart rate consistent across all levels of trust. Future research could include a deeper investigation in the relationship between trust and trusting behaviors.

It was somewhat surprising to discover that there was no difference between the control and high SA condition with regards to the time and distance to take over. One might wonder about the relative benefits associated with the high SA vs. the control condition with regards to the two behavioral measures. However, the results related to the monitoring ratio may help us better interpret our results relative to the benefits of the high SA condition. More specifically, participants had a lower monitoring ratio in the high SA condition than in the control condition. This indicates that while participants may have waited as long to take over the driving and reached similar distances before switching lanes, in the high SA condition participants were able to spend more time focusing on the task and less time monitoring the road. Thus, there are significant potential benefits to the high SA condition.

Fourth, this study contributes to the literature by examining the driver performance on a secondary task. Prior studies have treated secondary tasks as distractions to avoid. However, to fully leverage an automated driving system, the driver needs to be able to fully engage in another task while the automated vehicle is driving. Yet, we know very little about the factors that might encourage the driver to effectively engage in another task under these conditions. Future studies should be conducted along this avenue to further our understanding.

Finally, we found that the effectiveness of trust in automation on the secondary task performance depends on the level of situational awareness. Trust in automation alone did not appear to be sufficient to produce better performance. Simply put, trust is not enough. When participants trusted the automated vehicle and the vehicle supported their situational awareness, they were able to fully leverage the capabilities of the automated vehicle (see Figure 5). However, when participants trusted the automated vehicle and it did not project what needed to be done in the future, the automated vehicle trust was not associated with better secondary task performance.

This finding sheds new insights into when trusting the automated vehicle is important with regard to secondary task performance.

## 6. Limitations and Conclusion

This study has several limitations. Our experimental study used a simulated environment. Participants may act and react differently in an actual vehicle in a less controlled environment. Although there is some evidence which demonstrates that individuals respond similarly to both actual and simulated environments (Heydarian et al., 2015), we acknowledge that the use of a simulator and the experimental setting are both potential limitations. The age, gender and dominant hand of our participants may have also been a limitation. The participants' mean age was 25.7, 73.33% were male and 96.67% were right handed. Therefore, we should be cautious when attempting to generalize the results of this study to other demographics. Another limitation of the study is that the participants lacked prior experience working with advanced vehicle safety features. According to prior research (Yang et al., 2017), people's experience with an automated technology affects their trust in automation. To capture human drivers' change in trust and behaviors over time, a longitudinal study would be required.

In conclusion, drivers are failing to fully leverage the automated vehicle because of a lack of trust. This study provides evidence that driver assistance systems that support situational awareness can help alleviate this problem. Results of this study provide an interesting starting point. Yet, much more research is needed. We hope future research can build on the results of this study and expand our knowledge on this important topic.

### Acknowledgments

This research was supported in part by the Automotive Research Center (ARC) at the University of Michigan, with funding from government contract Department of the Army W56HZV-14-2-0001 through the U. S. Army Tank Automotive Research, Development, and Engineering Center (TARDEC). The authors acknowledge and greatly appreciate the guidance of Victor Paul (TARDEC), Ben Haynes (TARDEC), and Jason Metcalfe (ARL) in helping design the study. The authors would also like to thank Quantum Signal, LLC, for providing its ANVEL software and invaluable development support.